\documentclass[
aps,
prd,
tightenlines,
superscriptaddress,
nofootinbib,
floatfix]
{revtex4}
\usepackage{graphicx,
longtable
}

\begin{document}
\title{Global analysis of neutrino masses, mixings and phases:\\
 entering the era of leptonic CP violation searches}
\author{		G.L.~Fogli}
\affiliation{   	Dipartimento Interateneo di Fisica ``Michelangelo Merlin,'' 
               		Via Amendola 173, 70126 Bari, Italy}
\affiliation{   	Istituto Nazionale di Fisica Nucleare, Sezione di Bari, 
               		Via Orabona 4, 70126 Bari, Italy}
\author{		E.~Lisi}
\affiliation{   	Istituto Nazionale di Fisica Nucleare, Sezione di Bari, 
               		Via Orabona 4, 70126 Bari, Italy}
\author{		A.~Marrone}
\affiliation{   	Dipartimento Interateneo di Fisica ``Michelangelo Merlin,'' 
               		Via Amendola 173, 70126 Bari, Italy}
\affiliation{   	Istituto Nazionale di Fisica Nucleare, Sezione di Bari, 
               		Via Orabona 4, 70126 Bari, Italy}
\author{		D.~Montanino}
\affiliation{   	Dipartimento  di Matematica e Fisica ``Ennio De Giorgi,'' 
               		Via Arnesano, 73100 Lecce, Italy}
\affiliation{   	Istituto Nazionale di Fisica Nucleare, Sezione di Lecce, 
               		Via Arnesano, 73100 Lecce, Italy}
\author{		A.~Palazzo}
\affiliation{ 		Cluster of Excellence, Origin and Structure of the Universe, 
					  Technische Universit\"at M\"unchen,
					  Boltzmannstra\ss{e} 2, D-85748 Garching, Germany}
\author{		A.M.~Rotunno}
\affiliation{   	Dipartimento Interateneo di Fisica ``Michelangelo Merlin,'' 
               		Via Amendola 173, 70126 Bari, Italy}

\begin{abstract}
We perform a global analysis of neutrino oscillation data, including  
high-precision measurements of the neutrino mixing angle $\theta_{13}$ at reactor experiments,
which have confirmed previous indications in favor of $\theta_{13}>0$. Recent data presented at
the {\em Neutrino 2012\/} Conference are also included.  We focus
on the correlations
between $\theta_{13}$ and the mixing angle $\theta_{23}$, as well as between $\theta_{13}$ and the
neutrino CP-violation phase $\delta$. We find interesting indications
for $\theta_{23}< \pi/4$ and possible hints for $\delta\sim \pi$, with no
significant difference between normal and inverted mass hierarchy. 
  
\end{abstract}
\maketitle


\section{Introduction}

Current neutrino oscillation experiments (except for a few anomalous results) can be
interpreted in a simple three-neutrino framework, where the three flavor states  $\nu_\alpha=(\nu_e,\nu_\mu,\nu_\tau)$ 
are quantum superpositions of three light mass states $\nu_i=(\nu_1,\,\nu_3,\,\nu_3)$ 
via a unitary mixing matrix $U_{\alpha i}$, parametrized in terms
of three mixing angles $(\theta_{12},\theta_{13},\theta_{23})$
and one possible CP-violating phase $\delta$ in standard notation \cite{Na10,Nu12}. 

In neutrino oscillations, CP violation is a genuine $3\nu$ effect which may be 
observed (provided that $\delta\neq 0,\pi$) only  if all
the mixings $\theta_{ij}$ and the squared mass differences $m^2_i-m^2_j$ are nonzero \cite{Ca78}. 
The latter condition is experimentally established, and 
can be expressed in terms of the two independent parameters $\delta m^2=m^2_2-m^2_1>0$ \cite{Na10} and 
$\Delta m^2=m^2_3-(m^2_1+m^2_2)/2$ \cite{Fo06}, 
where $\Delta m^2>0$ ($<0$) corresponds to normal (inverted) mass spectrum hierarchy. 

Until very recently, the further condition $\theta_{ij}\neq 0$ could be considered as established for 
$\theta_{12}$ and $\theta_{23}$ \cite{Na10}, and quite likely (at $\sim 3\sigma$ level) but not conclusively
settled for $\theta_{13}$ \cite{Fo11}.  
This year, the short-baseline (SBL)  reactor experiments Daya Bay \cite{Daya} 
and RENO \cite{RENO} have definitely established that $\theta_{13}>0$ at $\sim 5\sigma$, by observing $\overline\nu_e$ disappearance
from near to far detectors. In particular, Daya Bay and RENO have measured $\sin^2\theta_{13}\simeq 0.023\pm0.003$ \cite{Day2}
and  $\sin^2\theta_{13}\simeq 0.029\pm0.006$ \cite{RENO,REN2}, respectively.  Consistent indications were also found
in the Double Chooz reactor experiment 
with far detector only ($\sin^2\theta_{13}\simeq 0.028\pm0.010$) \cite{DCho,DCh2}.
All these reactor data are in good agreement
with the results of our latest global analysis of oscillation data in \cite{Fo11}, which provided $\sin^2\theta_{13}=0.021$--0.025 at best fit,
with a $1\sigma$ error of $\pm 0.007$. 

It should be remarked that we had previously obtained hints in favor of $\sin^2\theta_{13}\sim 0.02$
from a detailed analysis of solar and long-baseline reactor data \cite{NOVE,HINT} (see also \cite{Baha} for similar, independent hints),
consistently with an earlier (weak) preference for $\theta_{13}>0$ from atmospheric neutrinos \cite{Fo06,HINT}. The hints 
became a $\sim\!\! 2\sigma$ indication for $\theta_{13}>0$ in combination with 
early appearance data from the MINOS long-baseline accelerator experiment \cite{Ve09}, and provided a 
$>3\sigma$ evidence by including the remarkable low-background appearance data from the T2K experiment \cite{Fo11}. 
The Daya Bay and RENO measurements have shown that our global $3\nu$ analyses in \cite{NOVE,HINT,Fo11}---the 
latest of a series started two decades ago \cite{COMP}---were on the 
right track in the hunt to $\theta_{13}$. See also \cite{Go10,Vall,Ma11}
for other recent analyses of $\theta_{13}$ constraints prior to the Daya Bay and RENO results.

With $\sin^2\theta_{13}$ as large as 2--3$\times 10^{-2}$, the door is open to CP violation searches in the neutrino sector, although
the road ahead appears to be long and difficult \cite{Via1,Via2}. At present, it makes sense to squeeze, from the available data, any tiny bit 
of information about $\delta$. An interesting attempt has been made in \cite{Mina}, using reactor and accelerator data. 
However, atmospheric $\nu$ data may also usefully probe $\delta$ \cite{Pe04,Fo06}. To this purpose, we update the analysis in \cite{Fo11}
by including new atmospheric, LBL accelerator and SBL reactor data, as available after the {\em Neutrino 2012\/} Conference
\cite{Nu12}. We have also extended
our atmospheric $\nu$ codes (previously limited to $\cos\delta=\pm 1$ \cite{Fo06,Fo11})
to generic values of $\delta$. Among the results obtained, we pay particular attention on a possible
preference in favor of $\theta_{23}<\pi/4$ and of $\delta\sim \pi$ in both
hierarchies (although with limited statistical significance). 
We also discuss the implications of the oscillation parameter constraints for absolute $\nu$ 
mass searches, as well as some limitations and challenges of global analyses. 

The present work is structured as follows. In Sec.~II we describe some methodological issues, 
which may be skipped by readers interested only in the main results. In Sec.~III we discuss
the results of our analysis in terms of covariance among the parameters $(\sin^2\theta_{13},\,\sin^2\theta_{23},\,\delta)$,
for both normal and inverted hierarchy. In Sec.~IV we summarize the
constraints on the mass-mixing oscillation parameters, and describe their
implications for the observables sensitive to absolute neutrino masses.
We conclude our work in Sec.~V. 
Details of atmospheric neutrino flavor evolution for
generic $\delta$ are confined in the Appendix.

\newpage
\section{Methodology: grouping and analyzing different data sets}
\vspace*{-1.5mm}

No single oscillation experiment can sensitively probe, at present, the full parameter space spanned
by $(\delta m^2,\,\pm\Delta m^2,\,\theta_{12},\,\theta_{13},\,\theta_{23},\,\delta)$. Therefore,
it is necessary to group in some way the experimental data, in order to study their
impact on the oscillation parameters. For instance, in \cite{Fo11} we showed that consistent indications
in favor of nonzero $\theta_{13}$ emerged from two different datasets, one
mainly sensitive to $\delta m^2$ (solar plus KamLAND experiments) and another mainly sensitive to $\Delta m^2$ 
(CHOOZ plus atmospheric and LBL accelerator experiments). In this work we adopt an alternative grouping of
datasets, which is more appropriate to discuss interesting features of the 
current data analysis, 
such as the covariance among the parameters $(\sin^2\theta_{13},\,\sin^2\theta_{23},\,\delta)$ 
in both mass hierarchies.

\vspace*{-4mm}
\subsection{LBL + solar + KamLAND data}
\vspace*{-1mm}

We remind that LBL accelerator data (from the K2K, T2K, and MINOS experiments) in the
$\nu_\mu\to\nu_\mu$ disappearance channel probe dominantly
the $\Delta m^2$-driven amplitude 
\begin{equation}
\label{mumu}
                         |U_{\mu3}|^2(1-|U_{\mu 3}|^2) = \cos^2\theta_{13}\sin^2\theta_{23}(1-\cos^2\theta_{13}\sin^2\theta_{23})\ ,  
\end{equation}
which is slightly octant-asymmetric in $\theta_{23}$ for $\theta_{13}\neq 0$. 
In the $\nu_\mu\to\nu_e$ appearance channel, the dominant $\Delta m^2$-driven amplitude  is 
\begin{equation}
\label{mue}
                         |U_{\mu3}|^2|U_{e 3}|^2 = \cos^2\theta_{13}\sin^2\theta_{13}\sin^2\theta_{23}\ ,  
\end{equation}
which is definitely octant-asymmetric in $\theta_{23}$ for $\theta_{13}\neq 0$. In both the appearance and the disappearance
channels, subdominant terms driven by $\delta m^2$ and by matter effects can also contribute to lift the octant symmetry 
and to provide some weak sensitivity to sign$(\Delta m^2)$ and to $\delta$, see e.g.~\cite{Asan} for a recent
perturbative approach at ``large'' $\theta_{13}$. As already noted in
\cite{Fo11}, the T2K and MINOS indications in favor of $\nu_\mu\to\nu_e$ appearance induce an anti-correlation, via Eq.~(\ref{mue}), between the preferred values
of $\sin^2\theta_{23}$ and $\sin^2\theta_{13}$. This covariance is relevant in the analysis of the $\theta_{23}$ octant
degeneracy \cite{Fo96}  and has an indirect impact also on the preferred ranges of $\delta$ via subdominant effects.

In order to make the best use of LBL accelerator data, it is thus useful to: (1) analyze both disappearance and
appearance data at the same time and in a full $3\nu$ approach; (2) 
combine LBL with solar and KamLAND data, which provide independent constraints on
$(\delta m^2,\theta_{12},\theta_{13})$ and thus on the subdominant $3\nu$ oscillation terms.
As discussed below, once the (relatively well known) oscillation parameters $\sin^2\theta_{12}$, $\delta m^2$ and $\Delta m^2$ are
marginalized away, interesting correlations emerge among the remaining parameters $(\sin^2\theta_{13},\,\sin^2\theta_{23},\,\delta)$. 
Conversely, these interesting bits of information are partly lost if LBL disappearance
data are analyzed in the $2\nu$ approximation and/or separately from appearance data, as it has often been the case in
official analyses by experimental collaborations. 

In this work, the previous LBL data used in \cite{Fo11} are updated with the inclusion of 
the first T2K disappearance constraints \cite{T2KD} and of the latest T2K appearance data \cite{T2K2}.
We note that recent MINOS $\overline\nu_\mu$ disappearance data \cite{MINA} are no longer in disagreement  with previous
$\nu_\mu$ results. Therefore, it makes sense to use both $\nu$ and $\overline\nu$ MINOS disappearance constraints, 
which we take from \cite{MIN2}, together with updated MINOS appearance data.
For later purposes, we note that recent T2K and (especially) MINOS data are best fit for slightly nonmaximal mixing 
($\sin^2 2\theta_{23}\simeq 0.94$--0.98 \cite{T2KD,MINA,MIN2}) roughly corresponding to the octant-symmetric values 
$\sin^2\theta_{23}\sim 0.4$ or 0.6).
A slight preference for nonmaximal mixing emerged also from our analysis of K2K LBL data in \cite{Fo06}.

\vspace*{-4mm}
\subsection{Adding SBL reactor data}
\vspace*{-1mm}

After grouping LBL accelerator plus solar plus KamLAND data (LBL + solar + KamLAND), 
it is important to add the independent and ``clean'' constraints on $\theta_{13}$ coming from SBL reactor experiments in
the $\nu_e\to\nu_e$ disappearance channel, which probe dominantly the $\Delta m^2$-driven amplitude 
\begin{equation}
\label{ee}
                         |U_{e3}|^2(1-|U_{e 3}|^2) = \sin^2\theta_{13}\cos^2\theta_{13}\ .  
\end{equation}
In the reactor dataset, subdominant terms are slightly sensitive to $(\delta m^2,\theta_{12})$ and, as noted
in \cite{Pa02} and discussed in \cite{Piai}, probe also the neutrino mass hierarchy.  
We include far-detector data from CHOOZ \cite{CHOO} and Double Chooz \cite{DCh2} and near-to-far detector constraints
from Daya Bay \cite{Day2} and RENO \cite{RENO,REN2}.  We do not include data from pre-CHOOZ reactor experiments, which
mainly affect normalization issues.
 
Indeed, the analysis of reactor experiments without near detectors depends, to some
extent, on the absolute normalization of the neutrino fluxes, which we choose 
to be the ``old'' (or ``low'') one, in the terminology of \cite{Fo11}.  
We shall also comment on the effect of adopting the ``new'' (or ``high'') normalization recently proposed in \cite{Ano1,Hube}. 
Constraints from Daya Bay and RENO are basically independent of such normalization, which is left free in the official analyses 
and is largely canceled by comparing near and far rates of events \cite{Daya,RENO}. At present, it is not possible
to reproduce, from published information, the official Daya Bay and RENO data analyses with the permill accuracy 
appropriate to deal with the small systematics affecting near/far ratios. We think that, for the purposes of this work, it is sufficient to
take their measurements of $\sin^2 2\theta_{13}$ at face value, as gaussian constraints on such parameter. 
Luckily, such constraints appear to depend very little on the  $\Delta m^2$ parameter within its currently allowed range; 
see the ($\Delta m^2,\sin^22 \theta_{13}$) prospective sensitivity plots in \cite{DaSe} (Daya Bay)
and \cite{RESe} (RENO). Of course, a joint analysis of all SBL reactor data made by the current collaborations would be desirable, since a few systematics
are correlated among the experiments.

As shown in \cite{Fo96}, LBL data in disappearance and appearance
mode  generally select [via Eqs.~(\ref{mumu}) and (\ref{mue})], 
two degenerate $(\theta_{23},\theta_{13})$  solutions,
characterized by nearly octant-symmetric values of $\theta_{23}$ and by slightly different values of $\theta_{13}$. By selecting a narrow range of $\theta_{13}$, 
precise reactor data can thus (partly) lift 
the $\theta_{23}$ octant degeneracy \cite{Fo96}  (see also \cite{AcRe}). Amusingly, the fit
results in Sec.~III resemble the
hypothetical, qualitative $3\nu$ scenario studied in \cite{Fo96}. 

\vspace*{-4mm}
\subsection{Atmospheric neutrino data}
\vspace*{-1mm}

After combining the (LBL + solar + KamLAND) and (SBL reactor) datasets, we finally add 
the Super-Kamiokande atmospheric neutrino data (SK atm.), as reported for the joint SK phases I--IV in \cite{Itow} (but with
no statistical $\nu/\overline\nu$ separation \cite{Itow}, which we cannot reproduce in detail). The SK data
span several decades in neutrino and antineutrino energy and pathlengths, both in vacuum and in matter, 
in all appearance and disappearance channels involving $\nu_\mu$  and $\nu_e$, and thus they embed an extremely
rich $3\nu$ oscillation physics. 

In practice, it is difficult to infer ---from atmospheric data--- clean $3\nu$ information beyond 
the dominant parameters $(\Delta m^2,\,\theta_{23})$. Subdominant oscillation effects 
are often smeared out over wide energy-angle spectra of events, and can be partly mimicked by
systematic effects. For this reason, ``hints'' coming from current atmospheric data should be taken with
a grain of salt, and should be possibly supported by independent datasets. 
For instance, we have attributed some importance to a weak preference for $\theta_{13}>0$ found 
from atmospheric SK data in \cite{Fo06}, only after it was independently supported by solar+KamLAND data \cite{HINT} and, later,
by LBL accelerator data \cite{Fo11}. Similarly, we have typically found a preference of atmospheric SK data 
for $\theta_{23}<\pi/4$ \cite{Fo06,Fo11}; in the next Section, we shall argue that such preference now finds some 
extra support in other datasets, and thus starts to be an interesting frontier to be explored.

The situation is more vague for $\delta$. We argued in \cite{Fo06} (and also found in \cite{Fo11})
that a slight electron excess in the atmospheric event spectra at sub-GeV
energies could be better fit with $\cos\delta = -1$ as compared with $\cos\delta = +1$, via 
interference terms \cite{Pe04,Fo06} in the oscillation probability. Since the analyses in \cite{Fo06,Fo11} were limited to 
the two CP-conserving cases $\cos\delta=\pm 1$,  
we have now extended our atmospheric neutrino codes to generic values of $\delta$ in the oscillation probability; 
details are given in the Appendix. We continue to find a preference for $\cos\delta\simeq -1$,
as described in the next Section. This possible hint for $\delta \sim \pi$ is roughly
consistent with the SK official (although preliminary) analyses in \cite{Itow,Nu10}, but is not clearly matched
by a similar hint coming from other data. This is another reason for choosing to present
atmospheric constraints only after the discussion of other datasets.
In conclusion, we think that is methodologically useful to show, in sequence, 
the impact of data from (LBL + solar + KamLAND), plus (SBL reactors), plus (SK atm.) 
experiments on the neutrino oscillation parameters. 

\vspace*{-4mm}
\subsection{Limitations and challenges of global analyses}
\vspace*{-1mm}

Our global analyses offer contributions to the discussion
on the neutrino oscillation phenomenology, but should not be considered as a substitute for the official 
oscillation analyses performed by the experimental collaborations, which include 
unpublished or unreproducible information.
Therefore, our estimated parameter ranges may be slightly offset with respect to those estimated by the 
collaborations in dedicated $3\nu$ data analyses (when available). Our educated guess is that possible 
offsets are $<1\sigma$ at present, and often much lower. Of course, even a fraction of one standard deviation may
matter when discussing hints at or below the $2\sigma$ level, as done in the next Section. However, the success
story of the indications of $\theta_{13}>0$ \cite{HINT,Fo11} shows that discussions of $\sim 2\sigma$ 
effects may still have some interest. 

Global $3\nu$ analyses will face several new challenges in the near future. As already remarked, a joint analysis of all reactor
data with near and far detectors (Daya Bay, RENO, Double Chooz) will be useful to get the most stringent 
constraints on $\theta_{13}$. The T2K and MINOS
long-baseline accelerator experiments are urged to abandon any $2\nu$  
approximation in the interpretation of their (disappearance) data, 
and focus on full-fledged $3\nu$ combinations of appearance plus disappearance data.
Increasing attention should be paid to refined features of the LBL analysis, such as the impact of cross section assumptions
on the oscillation parameter ranges \cite{Mart}. Future solar and long-baseline reactor data might slightly reduce  
the uncertainties of  the $(\theta_{12},\,\delta m^2)$ parameters, which drive subleading oscillation terms at higher energies.    
Concerning atmospheric $\nu$s and their associated systematics, we think that, while waiting for future
large-volume detectors and data, the existing SK atmospheric data have not yet exhausted their physics potential:
dedicated $3\nu$ analyses from the SK collaboration might reveal intriguing indications on $\theta_{23}$ and on $\delta$, 
especially if their Monte Carlo simulations were reprocessed by assuming full, unaveraged $3\nu$ oscillations from the very beginning
(rather than re-weighting unoscillated simulations with factors embedding averaged oscillations \cite{Wend}).

\newpage
\section{Results: correlations between $\theta_{13}$, $\theta_{23}$ and $\delta$}
\vspace*{-2mm}

In this section we focus on two emerging features of our analysis: converging
hints in favor of $\theta_{23}<\pi/4$, and a possible (weak) hint in favor of $\delta \sim \pi$. The correlations
of $\theta_{23}$ and $\delta$ with $\theta_{13}$ are discussed in some detail.  
As in our previous works \cite{Fo11,Fo06}, allowed regions are shown at $N\sigma$ confidence levels, 
where $N\sigma=\sqrt{\Delta\chi^2}$ \cite{Na10}. It is understood that, in each figure, undisplayed oscillation parameters have
been marginalized away. 

Figure~1 shows the results of the analysis in the plane ($\sin^2\theta_{13}, \,\sin^2\theta_{23}$), for both normal hierarchy (NH, upper panels)
and inverted hierarchy (IH, lower panels). From left to right, the panels
refer to increasingly rich datasets: LBL accelerator + solar + KamLAND data (left), plus SBL reactor data (middle), plus SK atmospheric data (right).

In the left panels, LBL appearance data anti-correlate  
$\sin^2\theta_{13}$  and $\sin^2\theta_{23}$ via Eq.~(\ref{mue}). On the other hand, LBL disappearance 
data (via their current preference for $\sin^2 2\theta_{23}<1$) disfavor maximal
mixing at $\gtrsim 1\sigma$. As a consequence, two quasi-degenerate $\chi^2$ minima emerge at complementary values  
of $\sin^2\theta_{23}$ and at somewhat different values of $\theta_{13}$.
The degeneracy is slightly lifted by solar+KamLAND data, whose
preference for $\sin^2\theta_{13}\simeq 0.02$ \cite{Fo11} picks up the first
octant solution in NH, and the second octant solution in IH. However, as far as LBL+solar+KamLAND data are concerned, 
the statistical difference between the 
two $\theta_{23}$ solutions remains negligible ($\lesssim 0.3\sigma$) in both NH and IH.

In the middle panels, the addition of SBL reactor data (most notably from Daya Bay and RENO) fixes $\sin^2\theta_{13}$ with high
accuracy and at relatively ``large'' values, which are best matched at low $\theta_{23}$---hence the overall preference
for the first $\theta_{23}$ octant in both hierarchies. Such preference is more pronounced in NH (at the level of $\sim 1\sigma$). In IH, 
both T2K and MINOS appearance data can accommodate values 
of $\theta_{13}$ generally larger than in NH \cite{TtoK,Mino,T2K2,MIN2} (as also evident from the left panels), so that the agreement with 
SBL reactor data can be easily reached in both octants, with only a small preference ($\sim\!0.4\sigma$) for the first. 
The combination of LBL accelerator and SBL reactor data to lift the octant
degeneracy was proposed in \cite{Fo96}. 

In the right panels, atmospheric $\nu$ data do not noticeably improve the constraints on $\theta_{13}$, but corroborate
the preference for the first octant (as already found in \cite{Fo06,Fo11}), in both
NH (slightly below the $3\sigma$ level) 
and IH (slightly below the $2\sigma$ level). [We do not observe an octant flip 
with the hierarchy as in \cite{Itow}.]
In conclusion, from Fig.~1 we derive that both atmospheric and non-atmospheric $\nu$ data seem to
prefer, independently, the first octant of $\theta_{23}$ (especially in normal hierarchy), 
with a combined statistical significance $\lesssim3\sigma$ in NH and $\lesssim 2\sigma$ in IH.
\vspace*{2mm}

\begin{figure}[b]
\includegraphics[width=0.82\textwidth]{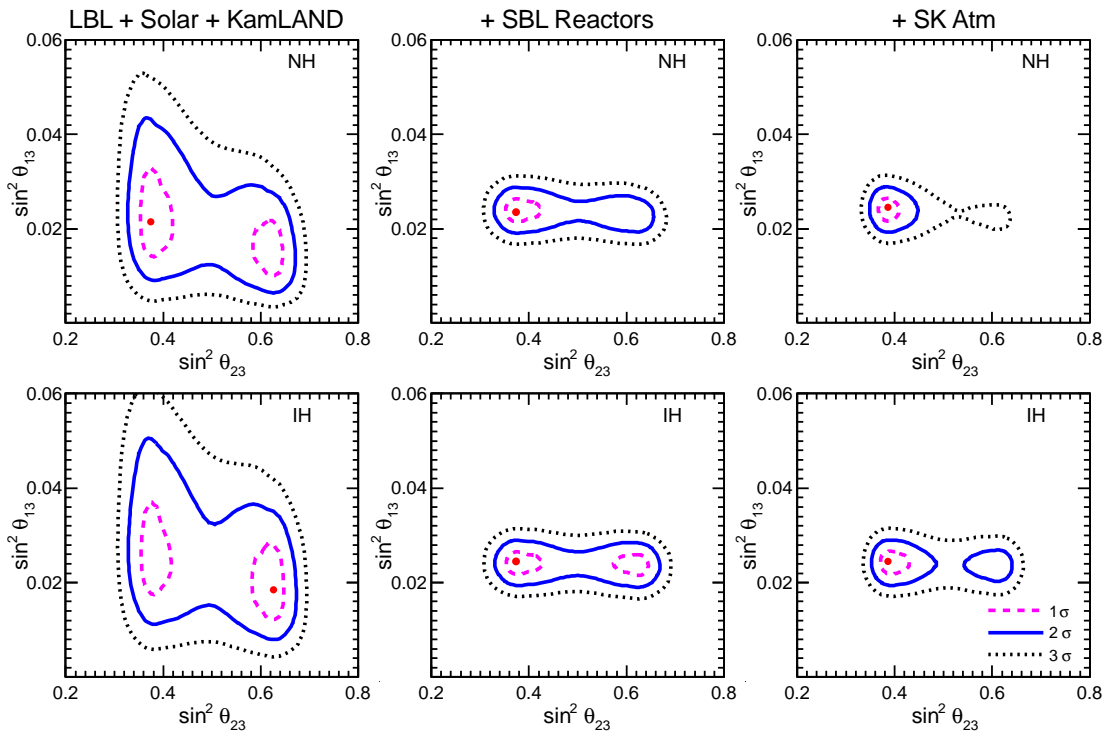}
\caption{\label{fig1} Results of the analysis in the plane charted by 
($\sin^2\theta_{13}, \,\sin^2\theta_{23}$), all other parameters being marginalized
away. From left to right, the 
regions allowed at 1, 2 and $3\sigma$ refer to increasingly rich datasets:
LBL+solar+KamLAND data (left panels), plus SBL reactor data (middle panels), plus SK atmospheric data (right panels).
Best fits are marked by dots.
A preference emerges for $\theta_{23}$ in the first octant in both normal hierarchy (NH, upper panels)
and inverted hierarchy (IH, lower panels).}
\end{figure}
\newpage

Figure~2 shows the results of the analysis in the plane ($\sin^2\theta_{13}, \,\delta$). The conventions used are the same as in Fig.~1.
Since the boundary values $\delta/\pi=0$ and $2$ are physically equivalent, each panel could be ideally ``curled'' by smoothly joining the
upper and lower boundaries.  

In the left panels, constraints on $\sin^2\theta_{13}$ are placed
both by solar+KamLAND data (independently of $\delta)$ and by current LBL accelerator data 
(somewhat sensitive to $\delta$). Once more, it can be noted that
larger values of $\theta_{13}$ are allowed in IH.
The best fit points are not statistically relevant, since
all values of $\delta$ provide almost equally good fits at $\sim 1\sigma$ level.  The ``fuzziness'' of the $1\sigma$ contours is 
a consequence of the statistical degeneracy of the two solutions allowed at $1\sigma$ in Fig.~1, and  
which involve complementary values of $\theta_{23}$ and somewhat different values of $\theta_{13}$.
At $1\sigma$, the fit is ``undecided'' between the wavy bands at smaller and larger values of $\theta_{13}$, and easily flips
between them. At $2$ or $3\sigma$ the two bands merge and such degeneracy effects are no longer apparent.  
  
In the middle panels, SBL reactor data pick up a very narrow range of $\theta_{13}$ and suppress
degeneracy effects. Some sensitivity to $\delta$ starts to emerge, since the ``wiggles'' of the
bands in the left panel best match the $\delta$-independent SBL reactor constraints on $\sin^2\theta_{13}$ only in certain ranges
of $\delta$. The match is generally easier in inverted hierarchy (where LBL data allow a larger $\theta_{13}$ range) than
normal hierarchy. 

In the right panels, atmospheric neutrino data induce a preference for $\delta \sim \pi$,
although all values of $\delta$ are still 
allowed at $\sim 2\sigma$.  Such a  preference is 
consistent with our previous analyses limited to $\cos\delta=\pm 1$ \cite{Fo06,Fo11}, 
where we found $\delta=\pi$ preferred over $\delta =0$,
in both normal and inverted hierarchy. As discussed in \cite{Fo06}, 
for $\delta\sim \pi$ the interference term in the oscillation probability
provide some extra electron appearance in the sub-GeV atmospheric neutrino data,
which helps fitting the slight excess of electron-like events in this sample. 
In our opinion, atmospheric data can provide valuable indications about the
phase $\delta$, which may warrant dedicated analyses by the SK 
experimental collaboration,
especially in combination with data from the T2K collaboration,
which uses SK as far detector and thus shares some systematics related
to final state reconstruction and analysis. 

Concerning the hierarchy, in the middle panels of Figs.~1 and 2 
(all data but SK atm.) we find a slight preference for IH with respect
to NH ($\Delta\chi^2\simeq -0.38$). The situation is reversed in the right panels
(all data, including SK atm.), where NH is slightly favored ($\Delta\chi^2\simeq +0.35$).
These fluctuations between NH and IH fits are statistically irrelevant.   
We conclude that, in our analysis of oscillation data, there are converging hints in favor
of $\theta_{23}<\pi/4$ (especially in NH), a possible hint in favor of $\delta\sim \pi$ (mainly from SK atm.\ data), and no hint
about the mass hierarchy.

\begin{figure}[b]
\includegraphics[width=0.82\textwidth]{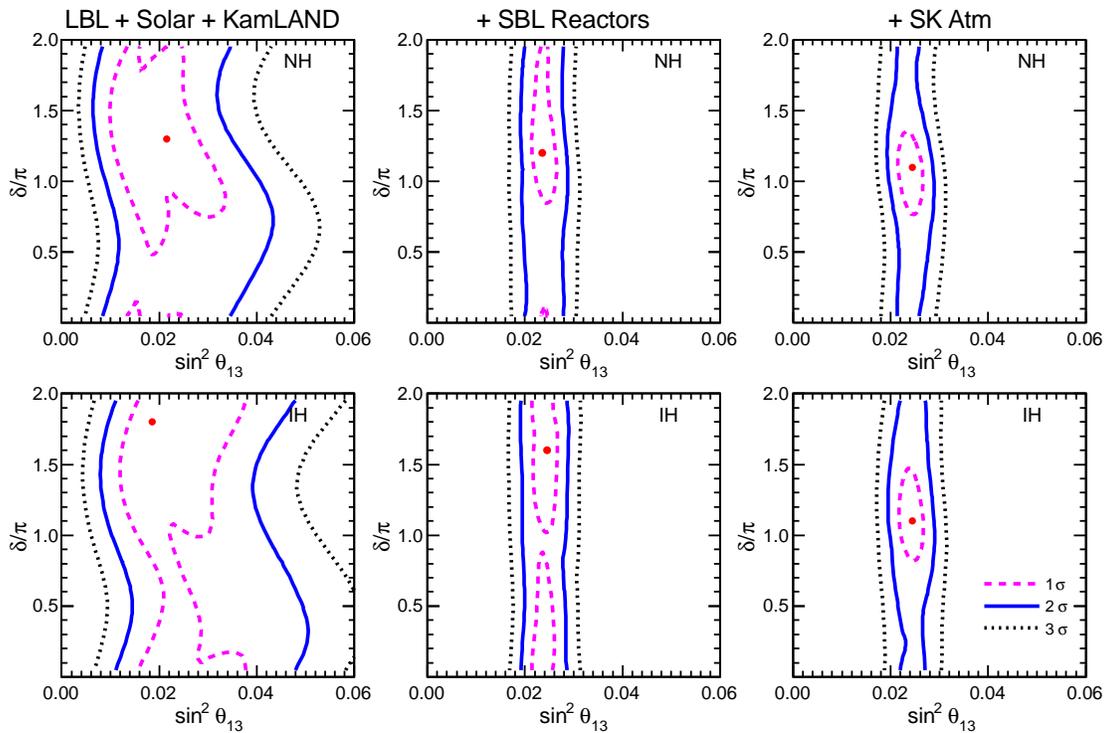}
\caption{\label{fig2} Results of the analysis in the plane charted by 
($\sin^2\theta_{13}, \,\delta$), all other parameters being marginalized
away. From left to right, the 
regions allowed at 1, 2 and $3\sigma$ refer to increasingly rich datasets:
LBL+solar+KamLAND data (left panels), plus SBL reactor data (middle panels), plus SK atmospheric data (right panels).
A preference emerges for $\delta$ values around $\pi$ in both normal hierarchy (NH, upper panels)
and inverted hierarchy (IH, lower panels).}
\end{figure}

\newpage
\section{Summary of oscillation constraints and implications for absolute masses}

In this section we summarize the previous results in terms of one-parameter constraints, all
the others being marginalized away. We also show updated oscillation constraints on the main absolute 
mass observables \cite{Obs1,Obs2}, namely, the effective electron neutrino mass $m_\beta$ (probed in $\beta$ decay),
the effective Majorana mass (probed in $0\nu2\beta$ decay searches), and the sum of neutrino masses
$\Sigma$, which can be probed by precision cosmology.

Figure~3 shows the $N\sigma$ bounds on the $3\nu$ oscillation parameters. Blue (solid) and red (dashed)
curves refer to NH and IH, respectively. The curves
are expected to be linear and symmetric around the best fit only for gaussian uncertainties.
This is nearly the case for the squared mass differences $\delta m^2$ and $\Delta m^2$, and for
the mixing parameters $\sin^2\theta_{12}$ and $\sin^2\theta_{13}$. The bounds on $\sin^2\theta_{23}$
are rather skewed towards the first octant, which is preferred at $\lesssim 2\sigma$ in NH
and $\lesssim 3\sigma$ in IH. Also 
the probability distribution of $\delta$ is highly nongaussian, with some preference for $\delta$ close to $\pi$,
but no constraint above $\sim\!2\sigma$. As expected, there are no visible differences between the NH and IH
curves for the parameters $\delta m^2$ and $\sin^2\theta_{12}$, and only minor variations for the the parameters
$\Delta m^2$ and $\sin^2\theta_{13}$. More pronounced (but $\lesssim 1\sigma)$ differences between NH and IH curves
can be seen for $\sin^2\theta_{23}$ and, to some extent, for $\delta$.

\begin{figure}[b]
\includegraphics[width=0.8\textwidth]{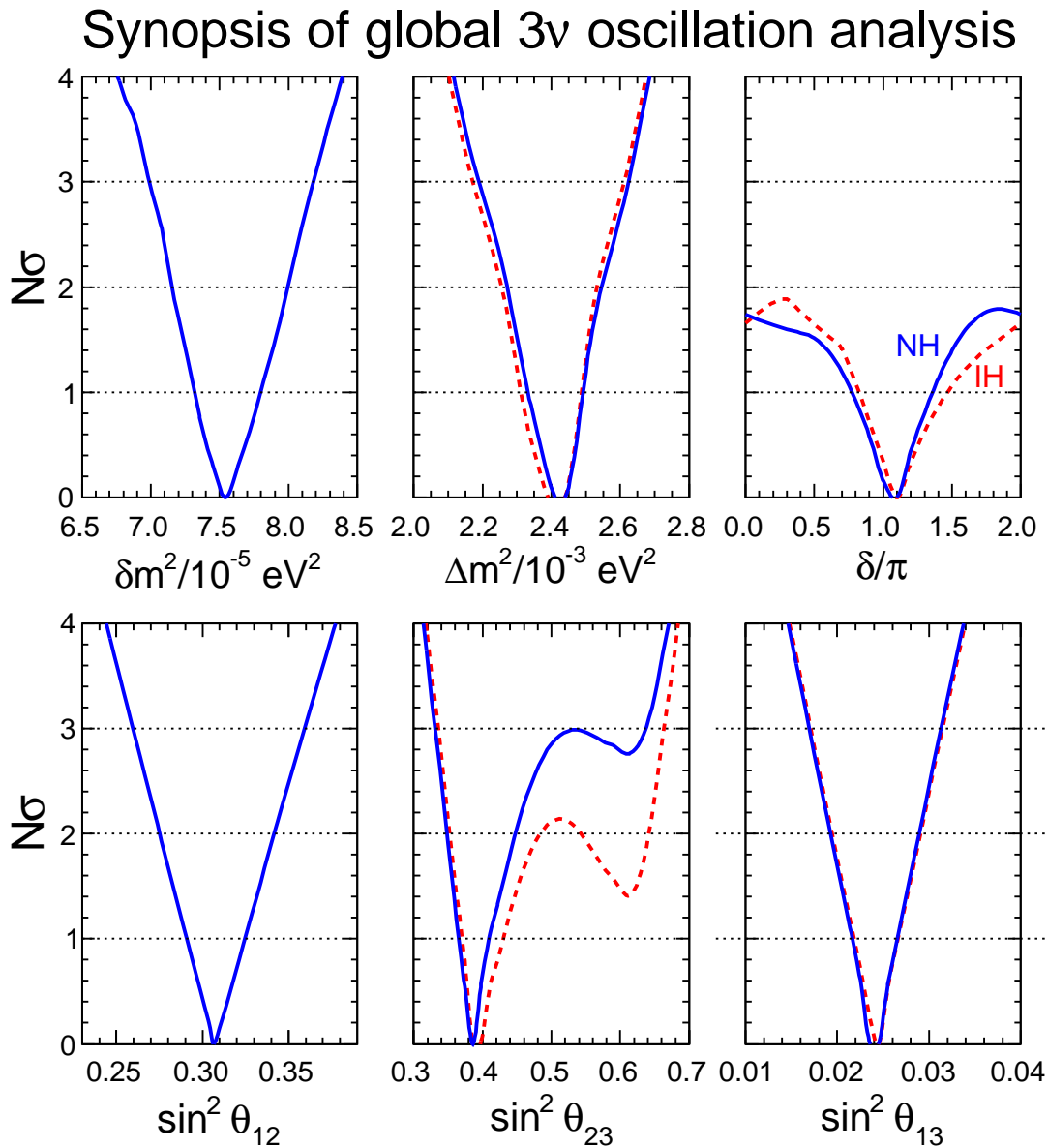}
\caption{\label{fig3} Results of the global analysis in terms of $N\sigma$
bounds on the six parameters governing $3\nu$ oscillations. Blue (solid) and red (dashed)
curves refer to NH and IH, respectively.}
\end{figure}

\begin{table}[t]
\caption{\label{Synopsis} Results of the global $3\nu$ oscillation analysis, in terms of best-fit values and
allowed 1, 2 and $3\sigma$ ranges  for the $3\nu$ mass-mixing parameters. We remind that $\Delta m^2$ is defined
herein as $m^2_3-{(m^2_1+m^2_2})/2$, with $+\Delta m^2$ for NH and $-\Delta m^2$ for IH.}
\begin{ruledtabular}
\begin{tabular}{lcccc}
Parameter & Best fit & $1\sigma$ range & $2\sigma$ range & $3\sigma$ range \\
\hline
$\delta m^2/10^{-5}~\mathrm{eV}^2 $ (NH or IH) & 7.54 & 7.32 -- 7.80 & 7.15 -- 8.00 & 6.99 -- 8.18 \\
\hline
$\sin^2 \theta_{12}/10^{-1}$ (NH or IH) & 3.07 & 2.91 -- 3.25 & 2.75 -- 3.42 & 2.59 -- 3.59 \\
\hline
$\Delta m^2/10^{-3}~\mathrm{eV}^2 $ (NH) & 2.43 & 2.33 -- 2.49 & 2.27 -- 2.55 & 2.19 -- 2.62 \\
$\Delta m^2/10^{-3}~\mathrm{eV}^2 $ (IH) & 2.42 & 2.31 -- 2.49 & 2.26 -- 2.53 & 2.17 -- 2.61 \\
\hline
$\sin^2 \theta_{13}/10^{-2}$ (NH) & 2.41 & 2.16 -- 2.66 & 1.93 -- 2.90 & 1.69 -- 3.13 \\
$\sin^2 \theta_{13}/10^{-2}$ (IH) & 2.44 & 2.19 -- 2.67 & 1.94 -- 2.91 & 1.71 -- 3.15 \\
\hline
$\sin^2 \theta_{23}/10^{-1}$ (NH) & 3.86 & 3.65 -- 4.10 & 3.48 -- 4.48 &                       3.31 -- 6.37 \\
$\sin^2 \theta_{23}/10^{-1}$ (IH) & 3.92 & 3.70 -- 4.31 & 3.53 -- 4.84 $\oplus$ 5.43 -- 6.41 & 3.35 -- 6.63 \\
\hline
$\delta/\pi$ (NH) & 1.08 & 0.77 -- 1.36 & --- & --- \\
$\delta/\pi$ (IH) & 1.09 & 0.83 -- 1.47 & --- & --- \\
\end{tabular}
\end{ruledtabular}
\end{table}

Table~I reports the bounds shown in Fig.~3 in numerical form. Except for $\delta$, the 
oscillation parameters are constrained with significant accuracy. If we define the average $1\sigma$
fractional accuracy as 1/6th of the $\pm 3\sigma$ variations around the best fit, then the 
parameters are globally determined with the following relative precision (in percent):
$\delta m^2$ (2.6\%), $\Delta m^2$ (3.0\%), $\sin^2\theta_{12}$ (5.4\%),
$\sin^2\theta_{13}$ (10\%), and $\sin^2\theta_{23}$ (14\%).

\begin{figure}[b]
\includegraphics[width=0.49\textwidth]{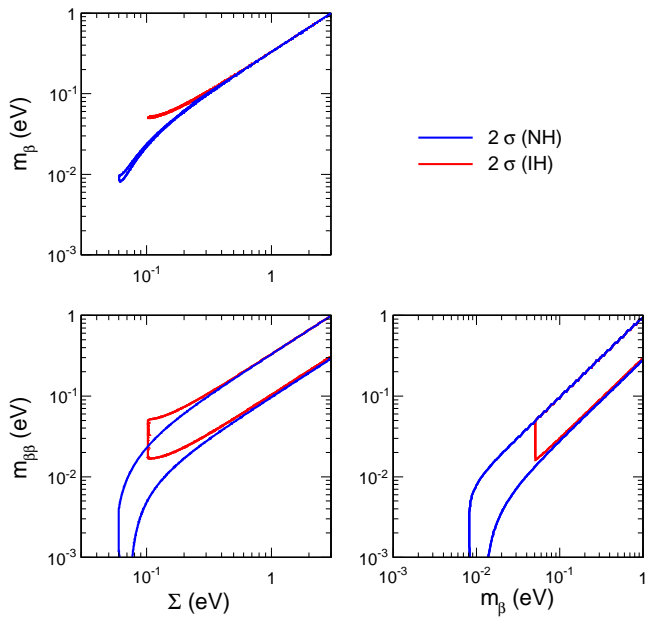}
\caption{\label{fig4} Constraints induced by oscillation data (at $2\sigma$ level) in the planes
charted by any two among the absolute mass observables $m_\beta$ (effective electron neutrino mass), $m_{\beta\beta}$
(effective Majorana mass), and $\Sigma$ (sum of neutrino masses).
Blue (red) bands refer to normal (inverted) hierarchy.}
\end{figure}

A final remark is in order. As noted in Sec.~II~B, two alternative choices were used in \cite{Fo11} for the
absolute reactor flux normalization, named as ``old'' and ``new,'' the latter being
motivated by revised flux calculations. Constraints were shown in \cite{Fo11} for both old and new normalization,
resulting in somewhat different values of $\theta_{12}$ and $\theta_{13}$. The precise
near/far data ratio constraints from Daya Bay \cite{Daya,Day2} and RENO \cite{RENO,REN2} 
are largely independent of
such normalization issues, which persists only for the reactor data without
near detector (i.e., KamLAND, CHOOZ and Double Chooz data in this work), with very small effects on the global fit.
For the sake of precision, we remark that the values in Table~I refer to our fit using the ``old'' normalization
for KamLAND, CHOOZ and Double Chooz. By using the ``new'' normalization, the only noticeable effects
would be the following overall shifts, with respect to the numbers in 
Table~I: $\Delta \sin^2\theta_{12}/10^{-1} \simeq +0.05$
and $\Delta \sin^2\theta_{13}/10^{-2}\simeq +0.08$ (i.e., at the level of $\sim 1/3$ of a standard
deviation).

\newpage
Let us now discuss the interplay of oscillation and nonoscillation data.
The constraints in  Table~\ref{Synopsis} induce strong covariances among the three main observables which
are sensitive to the absolute masses, namely, $m_\beta$, $m_{\beta\beta}$ and $\Sigma$ (see \cite{Obs1,Obs2} for notation). 
Figure~4 shows 
such covariances in terms of $2\sigma$ constraints (bands) in the planes charted by any couple of the 
absolute mass observables. As compared to previous results \cite{Obs1,Obs2}, the bands in the $(m_\beta,\,\Sigma)$ plane of Fig.~4 are narrower, 
due to the higher accuracy reached in the determination of all the oscillation parameters. Note that, in principle, 
precise measurements of $(m_\beta,\,\Sigma)$
in the sub-eV range (where the bands for NH and IH branch out) could determine the mass
spectrum hierarchy.
In the two lower panels of Fig.~4, 
there remains a large vertical spread in the allowed slanted bands, 
as a result of the unknown Majorana phases in the $m_{\beta\beta}$ components,
which may interfere either constructively (upper part of each band) or destructively (lower part of each band).
In principle, precise data in either the $(m_{\beta\beta},\,m_\beta)$ plane or the $(m_{\beta\beta},\,\Sigma)$ plane
might thus provide constraints on the Majorana phases.  

Progress in constraining the neutrino mass and mixing parameters will hopefully lead to a deeper understanding 
of their origin. Theoretical options range from ``accidental'' parameter values with no special 
significance or structure \cite{Acci} to ``special'' values pointing towards underlying symmetries \cite{Symm},
just to name a few possibilities in the vast literature on models. Precision measurements of neutrinos masses, mixings and phases
will provide valuable information to narrow this wide theoretical spectrum.

\section{Conclusions}

We have performed a global analysis of neutrino oscillation data, including recent, 
high-precision measurements of the neutrino mixing angle $\theta_{13}$ at reactor experiments
(which have confirmed previous indications in favor of $\theta_{13}>0$ \cite{HINT,Fo11}) and updated 
data released at the {\em Neutrino 2012\/} Conference \cite{Nu12}. 
We have explored the current correlations
between the mixing parameters $\sin^2\theta_{13}$ and $\sin^2\theta_{23}$, as well as between $\sin^2\theta_{13}$ and the
CP-violation phase $\delta$. We have found some interesting indications
in favor of $\theta_{23}< \pi/4$ (at $\lesssim 3\sigma$ in NH and $\lesssim 2\sigma$ in IH), as well as 
 possible hints of $\delta\sim \pi$, but
no significant difference between normal and inverted mass hierarchy. 
We surmise that full-fledged $3\nu$ analyses of LBL and atmospheric neutrino data
by the experimental collaborations would be very useful to better assess  the 
statistical relevance of these possible hints. 

\medskip\medskip
{\em Note added.} After this work was basically completed, we noted the results of another analysis 
including recent reactor data \cite{Tort}. Some differences with our results emerge in the
favored ranges for $\theta_{23}$ and $\delta$; they might be due, in part, to a different approach
to atmospheric neutrino oscillations 
(which, in our case, do include $\delta m^2$ and $\delta$ effects). We also
noted the preliminary results of the full $3\nu$ global analysis in \cite{Turn}, where $\theta_{23}<\pi/4$ is also preferred.

\acknowledgments
The authors acknowledge 
support by the Italian MIUR and INFN through the ``Astroparticle Physics'' 
research project. 
The work of A.P.\ is supported by the DFG Cluster of Excellence on the ``Origin and Structure of the Universe.''
Preliminary results of this work were presented by G.L.F.\ at the Workshop ``NuTurn 2012'' held at
INFN Laboratori Nazionali del Gran Sasso (Italy), at the Meeting ``European Strategy for Neutrino Oscillation Physics - II''
held at CERN, at the Meeting ``Rencontres de Blois 2012'' held in Blois (France), and at the Conference ``Neutrino 2012'' held in Kyoto (Japan).  


\appendix*
\section{Atmospheric neutrino flavor evolution for generic $\delta$}

Atmospheric neutrinos traverse the atmosphere and several Earth shells before being detected. We adopt a five-shell approximation of
the electron density $N$ in the Earth, in which each $j$-th cell  has sharp edge discontinuities and a mild dependence $N_j(r)$ in terms of 
the normalized radial distance $r$ from the Earth center~\cite{PREM}, that can be well approximated by a quartic polinomial~\cite{Lisi:1997yc}
\begin{equation}
	N_j(r) = \alpha_j + \beta_j r^2 + \gamma_j r^4
\ ,
\label{eq:Nj(r)}
\end{equation}
where the coefficients $\alpha_j$, $\beta_j$ and $\gamma_j$ are given in Table~I of~\cite{Lisi:1997yc}.

The evolution operator for atmospheric neutrinos can be written as the product of the evolution operator in each shell chord
\begin{equation}
{\cal T}_{\rm Earth}={\cal T}\left(\overline{P_0P_1}\right)\cdot{\cal T}\left(\overline{P_1P_2}\right)\cdot\ldots\cdot{\cal T}\left(\overline{P_{M-1}P_M}\right)\cdot{\cal T}_V\left(\overline{P_MP_A}\right)
\, ,\label{eq:Evolution}
\end{equation}
where $P_0$ is the detection point, $M$ the number of shells crossed by neutrinos and $P_A$ the production point in atmosphere. The last operator 
embeds the propagation in atmosphere, governed by the ``vacuum'' Hamiltonian ${\cal H}_v$. Notice that for a real Hamiltonian the calculation of ${\cal T}_{\rm Earth}$ can be further simplified using the symmetry properties of the electron density along the neutrino path inside the Earth (see appendix B of~\cite{Lisi:1997yc}). This property is no longer valid when the neutrino mixing matrix is not real, i.e., $\delta_{CP}\neq 0,\pi$.

A first-order 
approximation for the evolution operator inside the $k$-th shell is to consider the electron density constant, and equal to the
average along the shell chord
\begin{equation}
{\cal T}\left(\overline{P_{k-1}P_k}\right)=\exp\left[-i({\cal H}_v+{\overline V}_k)\cdot D_k\right]
\, ,\label{eq:Evol_shell}
\end{equation}
where ${\overline V}_k={\rm diag}\{\sqrt{2}G_F{\overline N}_k,0,0\}$ is the matter potential, 
$D_k$ the distance travelled by the neutrino inside the shell, and
\begin{equation}
{\overline N}_k=\frac{1}{D_k}\int_{x_{k-1}}^{x_k}dx\,N_k\left(\sqrt{x^2+\sin^2\eta}\right)
\, ,\label{eq:average}
\end{equation}
where 
\begin{equation}
\label{eta}
r^2=x^2+\sin^2\eta\, , 
\end{equation}
$\eta$ being the nadir angle of the neutrino direction. 
Handy subroutines for calculating exponentials of real or complex matrices can be found in the {\tt Expokit} package \cite{expokit}.
With the parameterization of Eq.~(\ref{eq:Nj(r)}), the integral in~Eq.~(\ref{eq:average}) is elementary.

A more accurate flavor evolution (beyond the constant-density approximation) 
can be obtained by applying the Magnus expansion~\cite{Magnus}, where the evolution operator is written as the exponential of an operator series, namely
\begin{equation}
{\cal T}(t)=\exp\left[\sum_{s=1}^\infty\Omega_s(t)\right]
\, ,\label{eq:Magnus}
\end{equation}
with
\begin{eqnarray}
\Omega_1(t) &=& -i\int_0^t dt_1\, {\cal H}_1\, ,\nonumber \\
\Omega_2(t) &=& -\frac{1}{2}\int_0^t dt_1\,\int_0^{t_1}dt_2\, 
\left[{\cal H}_1,{\cal H}_2\right]\, ,\nonumber \\
\Omega_3(t) &=& \frac{i}{6}\int_0^t dt_1\,\int_0^{t_1}dt_2\,\int_0^{t_2}dt_3\,\left(
\left[{\cal H}_1,\left[{\cal H}_2,{\cal H}_3\right]\right]+
\left[{\cal H}_3,\left[{\cal H}_2,{\cal H}_1\right]\right]
\right)
\, ,\label{eq:expansion}
\end{eqnarray}
and so on, where we have used the shorthand ${\cal H}_i\equiv{\cal H}(t_i)$. At first order, the Magnus expansion returns Eq.~(\ref{eq:Evol_shell}). 
At second order, it is $\left[{\cal H}_1,{\cal H}_2\right]=\left[{\cal H}_V,V(x_2)-V(x_1)\right]$. 
Integrating by part, one obtains
\begin{equation}
{\cal T}\left(\overline{P_{k-1}P_k}\right)=\exp\left[-i{\cal H}_k^{\rm eff}\cdot D_k\right]
\, ,\label{eq:Evol_shell1}
\end{equation}
with
\begin{equation}
{\cal H}_k^{\rm eff}={\cal H}_V+{\overline V}_k+i\left[{\cal H}_V,{\cal M}_k\right]
\, ,\label{eq:Heff}
\end{equation}
where
\begin{equation}
{\cal M}_k=\frac{1}{D_k}\int_{x_{k-1}}^{x_k}dx\, V(x)\left(x-\frac{x_{k-1}+x_k}{2}\right)
\, \label{eq:M_k}
\end{equation}
is the ``first moment'' of the matter potential around the trajectory midpoint 
inside the $k$-th shell. By using Eq.~(\ref{eta}) and the parameterization in Eq.~(\ref{eq:Nj(r)}),  
the integral in Eq.~(\ref{eq:M_k}) is elementary. 

Concerning the flavor evolution of atmospheric neutrinos, we have adopted the second-order Magnus expansion 
for generic (real or complex) Hamiltonian, and we have checked that this approximation
retains all the advantages of a fast analytical solution,
without introducing significant differences with respect to the more accurate (but slower) numerical 
integration along the Earth density profile. We have also checked that our codes reproduce well the oscillograms 
discussed in \cite{Gram} (not shown).

\newpage


\begin{thebibliography}{99}

\bibitem{Na10} 	K.~Nakamura and S.T.~Petcov, ``Neutrino mass, mixing, and oscillations,'' in
					J.\ Beringer {\em et al.} (Particle Data Group), Phys.\ Rev.\ D  {\bf 86}, 010001 (2012). 

\bibitem{Nu12} The status of neutrino oscillations has been recently reviewed in several presentations at {\em Neutrino 2012}, the 
	XXV International Conference on Neutrino Physics and Astrophysics (Kyoto, Japan, 2012),
	available at the website: neu2012.kek.jp~.


\bibitem{Ca78}   N.~Cabibbo,
  ``Time Reversal Violation in Neutrino Oscillation,''
  Phys.\ Lett.\ B {\bf 72}, 333 (1978).


\bibitem{Fo06}
 G.~L.~Fogli, E.~Lisi, A.~Marrone and A.~Palazzo,
  ``Global analysis of three-flavor neutrino masses and mixings,''
  Prog.\ Part.\ Nucl.\ Phys.\  {\bf 57}, 742 (2006)
  [arXiv:hep-ph/0506083].

\bibitem{Fo11} 
  G.~L.~Fogli, E.~Lisi, A.~Marrone, A.~Palazzo and A.~M.~Rotunno,
  ``Evidence of $\theta_{13}>0$ from global neutrino data analysis,''
  Phys.\ Rev.\ D {\bf 84}, 053007 (2011)
  [arXiv:1106.6028 [hep-ph]].

\bibitem{Daya} 
  F.~P.~An {\it et al.}  [Daya-Bay Collaboration],
  ``Observation of electron-antineutrino disappearance at Daya Bay,''
Phys.\ Rev.\ Lett.\  {\bf 108}, 171803 (2012)
  [arXiv:1203.1669 [hep-ex]].

\bibitem{RENO} J.~K.~Ahn {\it et al.}  [RENO Collaboration],
  ``Observation of Reactor Electron Antineutrino Disappearance in the RENO Experiment,''
  Phys.\ Rev.\ Lett.\  {\bf 108}, 191802 (2012)
  [arXiv:1204.0626 [hep-ex]].

\bibitem{Day2}  D.\ Dwyer [for the Daya-Bay Collaboration], talk at {\em Neutrino 2012} \protect\cite{Nu12}.


\bibitem{REN2}  S.-B.\ Kim [for the RENO Collaboration], talk at {\em Neutrino 2012} \protect\cite{Nu12}.

\bibitem{DCho}   Y.~Abe {\it et al.}  [Double Chooz Collaboration],
  ``Indication for the disappearance of reactor electron antineutrinos in the Double Chooz experiment,''
  Phys.\ Rev.\ Lett.\  {\bf 108}, 131801 (2012)
  [arXiv:1112.6353 [hep-ex]].

\bibitem{DCh2} M.\ Ishitsuka [for the Daya Bay Collaboration], talk at {\em Neutrino 2012} \protect\cite{Nu12}.

\bibitem{NOVE} G.L.\ Fogli, E.\ Lisi, A.\ Marrone, A.\ Palazzo, and A.M.~Rotunno, 
  ``What we (would like to) know about the neutrino mass,''
  in the Proceedings
  of {\em NO-VE 2008}, IV International Workshop on
``Neutrino Oscillations in Venice'' (Venice, Italy, April 15-18, 2008),
 edited by M.~Baldo Ceolin (University of Padova, Papergraf Editions, Padova, Italy, 2008), p.~21;
 also available at:
neutrino.pd.infn.it/NO-VE2008 [arXiv:0809.2936 [hep-ph]].

\bibitem{HINT}  G.~L.~Fogli, E.~Lisi, A.~Marrone, A.~Palazzo and A.~M.~Rotunno,
  ``Hints of $\theta_{13} > 0$ from global neutrino data analysis,''
  Phys.\ Rev.\ Lett.\  {\bf 101}, 141801 (2008)
  [arXiv:0806.2649 [hep-ph]]. 


\bibitem{Baha}
  A.~B.~Balantekin and D.~Yilmaz,
  ``Contrasting solar and reactor neutrinos with a non-zero value of $\theta_{13}$,''
  J.\ Phys.\ G {\bf 35}, 075007 (2008)
  [arXiv:0804.3345 [hep-ph]].

\bibitem{Ve09} G.~L.~Fogli, E.~Lisi, A.~Marrone, A.~Palazzo and A.~M.~Rotunno,
  ``SNO, KamLAND and neutrino oscillations: $\theta_{13}$,''
  in {\em NEUTEL 2009},  Proceedings of the 13th International Workshop on Neutrino Telescopes (Venice, Italy, 2009), 
  published by M.~Baldo Ceolin (University of Padova, Papergraf Editions, Padova, Italy), p.~81
  [arXiv:0905.3549 [hep-ph]].


\bibitem{COMP} 
  G.~L.~Fogli, E.~Lisi and D.~Montanino,
  ``A comprehensive analysis of solar, atmospheric, accelerator and reactor neutrino experiments in a hierarchical three generation scheme,''
  Phys.\ Rev.\ D {\bf 49}, 3626 (1994).
 
 
\bibitem{Go10} 
  M.~C.~Gonzalez-Garcia, M.~Maltoni and J.~Salvado,
  ``Updated global fit to three neutrino mixing: status of the hints of $\theta_{13} > 0$,''
  JHEP {\bf 1004}, 056 (2010)
  [arXiv:1001.4524 [hep-ph]].

\bibitem{Vall} 
  T.~Schwetz, M.~Tortola and J.~W.~F.~Valle,
  ``Where we are on $\theta_{13}$: addendum to 'Global neutrino data and recent reactor fluxes: status of three-flavour oscillation parameters',''
  New J.\ Phys.\  {\bf 13}, 109401 (2011)
  [arXiv:1108.1376 [hep-ph]].

\bibitem{Ma11}
	M.\ Maltoni,
	``Status of three-neutrino oscillations,''
	Proceedings of {\em EPS-HEP 2011}, Europhysics Conference on High Energy Physics, 
	(Grenoble, France, 2011), published in PoS (EPS-HEP2011) 090, 2011.

\bibitem{Via1}
  P.~Huber, M.~Lindner, T.~Schwetz and W.~Winter,
  ``First hint for CP violation in neutrino oscillations from upcoming superbeam and reactor experiments,''
  JHEP {\bf 0911}, 044 (2009)
  [arXiv:0907.1896 [hep-ph]].

\bibitem{Via2}
  P.~Coloma, A.~Donini, E.~Fernandez-Martinez and P.~Hernandez,
  ``Precision on leptonic mixing parameters at future neutrino oscillation experiments,''
  arXiv:1203.5651 [hep-ph].

\bibitem{Mina}
 P.~A.~N.~Machado, H.~Minakata, H.~Nunokawa and R.~Z.~Funchal,
  ``Combining Accelerator and Reactor Measurements of $\theta_{13}$: The First Result,''
  JHEP {\bf 1205}, 023 (2012)
  [arXiv:1111.3330 [hep-ph]]. See also the earlier paper by 
  H.~Minakata and H.~Sugiyama,
  ``Exploring leptonic CP violation by reactor and neutrino superbeam experiments,''
  Phys.\ Lett.\ B {\bf 580}, 216 (2004)
  [hep-ph/0309323].

\bibitem{Pe04}  O.~L.~G.~Peres and A.~Yu.~Smirnov,
  ``Atmospheric neutrinos: LMA oscillations, $U_{e3}$ induced interference and CP violation,''
  Nucl.\ Phys.\ B {\bf 680}, 479 (2004)
  [hep-ph/0309312].

\bibitem{Asan}
  K.~Asano and H.~Minakata,
  ``Large-$\theta_{13}$ Perturbation Theory of Neutrino Oscillation for Long-Baseline Experiments,''
  JHEP {\bf 1106}, 022 (2011)
  [arXiv:1103.4387 [hep-ph]].

\bibitem{Fo96} 
  G.~L.~Fogli and E.~Lisi,
  ``Tests of three flavor mixing in long baseline neutrino oscillation experiments,''
  Phys.\ Rev.\ D {\bf 54}, 3667 (1996)
  [hep-ph/9604415].




\bibitem{T2KD} 
  K.~Abe {\it et al.}  [T2K Collaboration],
  ``First Muon-Neutrino Disappearance Study with an Off-Axis Beam,''
  Phys.\ Rev.\ D {\bf 85}, 031103 (2012)
  [arXiv:1201.1386 [hep-ex]].

\bibitem{T2K2} T.~Nakaya [for the T2K Collaboration], talk at {\em Neutrino 2012} \protect\cite{Nu12}.


\bibitem{MINA} 
  P.~Adamson {\it et al.}  [MINOS Collaboration],
  ``An improved measurement of muon antineutrino disappearance in MINOS,''
  arXiv:1202.2772 [hep-ex].
 

\bibitem{MIN2} R.~Nichol [for the MINOS Collaboration], talk at {\em Neutrino 2012} \protect\cite{Nu12}.
  
 
\bibitem{Pa02} 
 G.~L.~Fogli, E.~Lisi and A.~Palazzo,
  ``Quasi energy independent solar neutrino transitions,''
  Phys.\ Rev.\ D {\bf 65}, 073019 (2002)
  [hep-ph/0105080].
 
\bibitem{Piai}
   S.~T.~Petcov and M.~Piai,
  ``The LMA MSW solution of the solar neutrino problem, inverted neutrino mass hierarchy and reactor neutrino experiments,''
  Phys.\ Lett.\ B {\bf 533}, 94 (2002)
  [hep-ph/0112074].

 \bibitem{CHOO}
  M.~Apollonio {\it et al.}  [CHOOZ Collaboration],
  ``Search for neutrino oscillations on a long base-line at the CHOOZ  nuclear
  power station,''
  Eur.\ Phys.\ J.\  C {\bf 27}, 331 (2003)
  [arXiv:hep-ex/0301017].

 \bibitem{Ano1}  
  T.~A.~Mueller {\it et al.},
  ``Improved Predictions of Reactor Antineutrino Spectra,''
  Phys.\ Rev.\  C {\bf 83}, 054615 (2011)
  [arXiv:1101.2663 [hep-ex]].



\bibitem{Hube} P.~Huber,
 ``On the determination of anti-neutrino spectra from nuclear reactors,''
  Phys.\ Rev.\ C {\bf 84}, 024617 (2011)
  [Erratum-ibid.\ C {\bf 85}, 029901 (2012)]
  [arXiv:1106.0687 [hep-ph]].

 
\bibitem{DaSe}  X.~Guo {\it et al.}  [Daya-Bay Collaboration],
  ``A Precision measurement of the neutrino mixing angle theta(13) using reactor antineutrinos at Daya-Bay,''
  hep-ex/0701029.

\bibitem{RESe}   J.~K.~Ahn {\it et al.}  [RENO Collaboration],
  ``RENO: An Experiment for Neutrino Oscillation Parameter $\theta_{13}$ Using Reactor Neutrinos at Yonggwang,''
  arXiv:1003.1391 [hep-ex].



 \bibitem{AcRe} 
  K.~Hiraide, H.~Minakata, T.~Nakaya, H.~Nunokawa, H.~Sugiyama, W.~J.~C.~Teves and R.~Z.~Funchal,
  ``Resolving $\theta_{23}$ degeneracy by accelerator and reactor neutrino oscillation experiments,''
  Phys.\ Rev.\ D {\bf 73}, 093008 (2006)
  [hep-ph/0601258].


\bibitem{Itow} Y.~Itow, talk at \protect\cite{Nu12}.

  
\bibitem{Nu10} Y.\ Takeuchi [Super-Kamiokande Collaboration], 
 ``Results from Super-Kamiokande,'' in the Proceedings of {\em Neutrino 2010}, 
   XXIV International Conference on Neutrino Physics and Astrophysics
  (Athens, Greece, 2010), to appear. Website: http://www.neutrino2010.gr
  
\bibitem{Mart}  
    D.~Meloni and M.~Martini,
  ``Revisiting the T2K data using different models for the neutrino-nucleus cross sections,''
  arXiv:1203.3335 [hep-ph].

\bibitem{Wend}
R.~A.\ Wendell, ``Three flavor oscillation analysis of atmospheric neutrinos in Super-Kamiokande,''
                PhD Thesis, Duke University, 2008; available at www-sk.icrr.u-tokyo.ac.jp/sk/pub/
 
\bibitem{TtoK}
  K.~Abe {\it et al.}  [T2K Collaboration],
  ``Indication of Electron Neutrino Appearance from an Accelerator-produced
  Off-axis Muon Neutrino Beam,''
Phys.\ Rev.\ Lett.\  {\bf 107}, 041801 (2011)
  [arXiv:1106.2822 [hep-ex]].
 
\bibitem{Mino} 
 P.~Adamson {\it et al.}  [MINOS Collaboration],
  ``Improved search for muon-neutrino to electron-neutrino oscillations in MINOS,''
  Phys.\ Rev.\ Lett.\  {\bf 107}, 181802 (2011)
  [arXiv:1108.0015 [hep-ex]]. 
 
 

\bibitem{Obs1}  G.~L.~Fogli, E.~Lisi, A.~Marrone, A.~Melchiorri, A.~Palazzo, P.~Serra, J.~Silk and A.~Slosar,
  ``Observables sensitive to absolute neutrino masses: A Reappraisal after WMAP-3y and first MINOS results,''
  Phys.\ Rev.\ D {\bf 75}, 053001 (2007)
  [hep-ph/0608060].
 

\bibitem{Obs2}  G.~L.~Fogli, E.~Lisi, A.~Marrone, A.~Melchiorri, A.~Palazzo, A.~M.~Rotunno, P.~Serra and J.~Silk {\it et al.},
  ``Observables sensitive to absolute neutrino masses. II.,''
  Phys.\ Rev.\ D {\bf 78}, 033010 (2008)
  [arXiv:0805.2517 [hep-ph]].
 
 
\bibitem{Acci}  See, e.g., F.~Vissani,
  ``Expected properties of massive neutrinos for mass matrices with a dominant block and random coefficients order unity,''
  Phys.\ Lett.\ B {\bf 508}, 79 (2001)
  [hep-ph/0102236];
  J.~Gluza and R.~Szafron,
  ``Real and complex random neutrino mass matrices and theta13,''
  Phys.\ Rev.\ D {\bf 85}, 047701 (2012)
  [arXiv:1111.7278 [hep-ph]];
  A.~de Gouvea and H.~Murayama,
  ``Neutrino Mixing Anarchy: Alive and Kicking,''
  arXiv:1204.1249 [hep-ph].
 

\bibitem{Symm} See, e.g., G.~Altarelli and F.~Feruglio,
  ``Discrete Flavor Symmetries and Models of Neutrino Mixing,''
  Rev.\ Mod.\ Phys.\  {\bf 82}, 2701 (2010)
  [arXiv:1002.0211 [hep-ph]].

 
\bibitem{Tort}   D.~V.~Forero, M.~Tortola and J.~W.~F.~Valle,
  ``Global status of neutrino oscillation parameters after recent reactor measurements,''
  arXiv:1205.4018 [hep-ph].

\bibitem{Turn} T.~Schwetz,  talk at {\em NuTURN 2012}, Workshop on ``Neutrino at the Turning Point''
(Laboratori Nazionali del Gran Sasso, Italy, 2012), available at 
agenda.infn.it/conferenceDisplay.py?confId=4722; talk at {\em What is $\nu$?}, Workshop at the
Galileo Galilei Institute (Florence, Italy, 2012); available at www.ggi.fi.infn.it~.

\bibitem{PREM}	
  A.~M.~Dziewonski and D.~L.~Anderson,
  ``Preliminary Earth Model (PREM)''
  Phys.\ Earth Planet.\ Inter.\ {\bf 25}, 297 (1981).

\bibitem{Lisi:1997yc}
E.~Lisi and D.~Montanino,
  ``Earth regeneration effect in solar neutrino oscillations: An Analytic approach,''
  Phys.\ Rev.\ D {\bf 56}, 1792 (1997)
  [hep-ph/9702343].  
  
\bibitem{expokit}
 R.~B.~Sidje, ``Expokit: A Software Package for Computing Matrix Exponentials,''
 ACM Trans.\ Math.\ Softw.\ {\bf 24}, 130 (1998). Software package available at 
 www.maths.uq.edu.au/expokit/
  
\bibitem{Magnus} W.\ Magnus,
 ``On the exponential solution of differential equations for a linear operator,''
  Commun.\ Pure Appl.\ Math.\ {\bf 7}, 649 (1954).

\bibitem{Gram}  E.~Kh.~Akhmedov, M.~Maltoni and A.~Yu.~Smirnov,
  ``Neutrino oscillograms of the Earth: Effects of 1-2 mixing and CP-violation,''
  JHEP {\bf 0806}, 072 (2008)
  [arXiv:0804.1466 [hep-ph]].
 
  
\end{thebibliography}
\end{document}